\documentclass[12pt,english]{article}
\usepackage[T1]{fontenc}
\usepackage[latin9]{inputenc}
\usepackage{geometry}
\usepackage{color}
\usepackage{babel}
\usepackage{array}
\usepackage{float}
\usepackage{amsbsy}
\usepackage{amstext}
\usepackage{graphicx}
\usepackage{setspace}
\usepackage{esint}
\usepackage[authoryear]{natbib}
\usepackage[unicode=true,pdfusetitle,
 bookmarks=true,bookmarksnumbered=false,bookmarksopen=false,
 breaklinks=false,pdfborder={0 0 0},pdfborderstyle={},backref=false,colorlinks=false]
 {hyperref}

\makeatletter

\providecommand{\tabularnewline}{\\}

\usepackage{setspace}
\usepackage{lineno}
\usepackage{natbib}
\usepackage{pdflscape}
\usepackage{hyperref}
\usepackage{multirow}
\usepackage{amssymb}
\usepackage{xcolor}

\g@addto@macro\normalsize{%
  \setlength\abovedisplayskip{1pt}
  \setlength\belowdisplayskip{4pt}
  \setlength\abovedisplayshortskip{2pt}
  \setlength\belowdisplayshortskip{2pt}
}
\usepackage{titlesec}
\titlespacing*{\section}{0pt}{3ex plus 2ex}{1ex}
\titleformat*{\section}{\fontsize{14}{14}\bfseries}

\titlespacing*{\subsection}{0pt}{3ex plus 2ex}{1ex}
\titleformat*{\subsection}{\fontsize{12}{12}\bfseries}

\newcommand{\mycirc}[1][black]{\textcolor{#1}{\ensuremath\bullet}}

\makeatother

\begin{document}
\begin{flushleft}
\setcounter{footnote}{0}
\begin{flushleft}Running Head: Location uncertainty
\par\end{flushleft}

\noindent \textbf{Accounting for location uncertainty in distance
sampling data}

\medskip{}

\begin{singlespace}
\textbf{Trevor J. Hefley}

{\small{}Department of Statistics}{\small\par}

{\small{}Kansas State University}{\small\par}
\end{singlespace}

\begin{doublespace}
\medskip{}

\end{doublespace}

\begin{singlespace}
\textbf{W. Alice Boyle }

{\small{}Division of Biology}{\small\par}

{\small{}Kansas State University}{\small\par}
\end{singlespace}

\begin{doublespace}
\medskip{}

\end{doublespace}

\begin{singlespace}
\textbf{Narmadha M. Mohankumar}

{\small{}Department of Statistics}{\small\par}

{\small{}Kansas State University}{\small\par}
\end{singlespace}

\bigskip{}

\begin{singlespace}
\textbf{Statement of authorship: }T.J.H and W.A.B conceived the study.
T.J.H developed the statistical methods. T.J.H and N.M.M applied the
statistical methods, conducted the simulation experiment and data
analysis. T.J.H wrote the manuscript. All authors contributed substantially
to revisions. \bigskip{}

\textbf{Data accessibility statement:} The field-collected avian data
is publicly available from \citet{KONZADATA}. Additional files required
to reproduce the results of this study (e.g. shapefiles of transects)
will be archived in the Dryad Digital Repository.

\bigskip{}

\textbf{Reproducibility statement:} Annotated computer code capable
of reproducing all results and figures associated with the simulation
experiment and data analysis are provided in Appendix S2 and S3.\bigskip{}

\textbf{Article type: }Research article

\textbf{Word count:} 269 (abstract) 6899 (total) 1305 (references)

\textbf{Number of references: }54

\textbf{Number of figures: }3

\textbf{Number of tables: }2\medskip{}

\end{singlespace}

\setlength{\parindent}{0.7cm}

\pagebreak{}

\section*{Summary}
\begin{enumerate}
\item Ecologists use distance sampling to estimate the abundance of plants
and animals while correcting for undetected individuals. By design,
data collection is simplified by requiring only the distances from
a transect to the detected individuals be recorded. Compared to traditional
design-based methods that require restrictive assumption and limit
the use of distance sampling data, model-based approaches enable broader
applications such as spatial prediction, inferring species-habitat
relationships, unbiased estimation from preferentially sampled transects,
and integration into multi-type data models. Unfortunately, model-based
approaches require the exact location of each detected individual
in order to incorporate environmental and habitat characteristics
as predictor variables.
\item Using a hierarchical specification, we modified model-based methods
for distance sampling data by including a probability distribution
that accounts for location uncertainty generated when only the distances
are recorded. We tested and demonstrated our method using a simulation
experiment and by modeling the habitat use of Dickcissels (\textit{Spiza
americana}) using distance sampling data collected from the Konza
Prairie in Kansas, USA.
\item Our results showed that ignoring location uncertainty can result in
biased coefficient estimates and predictions. However, accounting
for location uncertainty remedies the issue and results in reliable
inference and prediction.
\item Location uncertainty is difficult to eliminate when collecting some
types of ecological data. Like other types of measurement error, hierarchical
models can accommodate the data collection process thereby enabling
reliable inference. Our approach is a significant advancement for
the analysis of distance sampling data because it remedies the deleterious
effects of location uncertainty and requires only distances be recorded.
In turn, this enables historical distance sampling data sets to be
compatible with modern data collection and modeling practices.\bigskip{}
\end{enumerate}
\begin{doublespace}
\noindent \textbf{\textit{\emph{Key-words: }}}ecological fallacy,
hierarchical model, integrated population model, point process, resource
selection, species distribution model \vspace{-1cm}

\end{doublespace}

\section*{Introduction}

Distance sampling has been widely used for nearly a half a century
to estimate abundance of plants and animals. This method involves
one or more observers recording the distances from point or line transects
to detected individuals (\citealt{burnham1980estimation,buckland2001introduction}).
Early statistical methods for the analysis of distance sampling data
used design-based estimators that accounted for errors in detection.
This resulted in a hybrid analysis, involving model-based methods
used to account for errors in detection and design-based methods used
to estimate abundance (\citealt{buckland2016model}). More recently,
fully model-based approaches have been developed to enable spatial
prediction, statistical inference regarding species-habitat relationships,
and unbiased estimation of abundance from point and line transects
that are placed non-randomly (\citealt{stoyan1982remark,hogmander1991random,hedley2004spatial,johnson2010model,miller2013spatial}).
Current areas of research include data assimilation, fusion, integration
or reconciliation requiring the development of joint models that combine
multiple types of data. Such recent developments include integrated
species distribution models that incorporate distance sampling data
and presence-only data collected by citizen scientists (\citeauthor{Fletcher2019}
\textit{in press}).

Spatially-explicit models that link abundance to environmental and
habitat characteristics are used in many areas of ecological research.
For example, presence-absence, count, and presence-only data enable
spatial prediction of species distributions (\citealt{elith2009species,hefleyHSDM}).
Other examples include integrated species distributions models used
to predict abundance and occupancy with higher accuracy by combining
multiple types of data (e.g., \citealt{williams2017integrated}; \citeauthor{Fletcher2019}
\textit{in press}). A common theme among these approaches is that
the location of the individual is conceptualized as a point in geographic
space where environmental conditions and habitat characteristics are
measured (\citealt{hefleyHSDM,kery2015applied,milller2019}). Those
location-specific conditions and characteristics (hereafter ``predictor
variables'') are used to specify an intensity function which enables
statistical inference regarding species-habitat relationships and
provides estimates of abundance and occupancy that can be mapped at
any spatial resolution. This framework relies on characterizing the
distribution of abundance as a spatial point processes which is the
same approach used to develop models for distance sampling data (\citealt{stoyan1982remark,hogmander1991random,hedley2004spatial,johnson2010model,miller2013spatial}). 

Often ecologists do not have the exact locations of individuals. For
example, exact locations are unrecoverable from distance sampling
data collected along line transects unless auxiliary information such
as the location of the observer at the time of detection is recorded.
Regardless of the mechanisms that obscure the exact locations, uncertainty
limits the usefulness of the data because values of the predictor
variables cannot be obtained. For example, if a distance sampling
data set does not include the exact locations then analysis is restricted
to models that include only the predictor variables that are constant
for all individuals detected from a given transect (\citealt{johnson2010model,buckland2016model}).
In practice, this lead to model-misspecification and lower predictive
accuracy. Sometimes researchers attempt to mitigate this problem by
using surrogate predictor variables such as the habitat characteristics
at a convenient location (e.g., the center of the transect) or the
average value of the predictor variable calculated from an area within
an arbitrary distance of the transect line or point. Use of surrogate
predictor variables can also bias parameter estimates and predictions.
In some cases, the bias can invert the inferred relationship between
predictor variables and abundance (\citealt{hefley2014correction,hefley2017bias,walker2018bias};
\citeauthor{walker2019bias} \textit{under revision}).

To eliminate these issues, we present a model-based approach for distance
sampling data that can be used when the location of individuals is
uncertain. Our approach enables the same inference as model-based
approaches requiring exact locations and can be incorporated into
integrated data models that are based on an underlying point process
(e.g., \citeauthor{Fletcher2019} \textit{in press}). Our approach
relies on a hierarchical modeling framework, but results in relatively
simple marginalized distributions that can be used for efficient Bayesian
or likelihood-based estimation and inference. Using simulated data,
we evaluate the ability of our method to account for location uncertainty
and compare it to existing approaches commonly used in practice. Finally,
we demonstrate our method using line transect data to estimate habitat
use of a grassland bird species, the Dickcissel. 
\begin{spacing}{1.9}

\section*{Materials and methods}
\end{spacing}
\begin{spacing}{1.9}

\subsection*{MODEL-BASED DISTANCE SAMPLING}
\end{spacing}

\noindent A common practice when constructing statistical models is
to choose a probability distribution that matches important characteristics
of the data. For example, if the data are counts then a statistical
model that assumes a Poisson distribution might be used. Counts must
be non-negative integers and the Poisson distribution is capable of
generating non-negative integers (i.e., the support of the data and
probability distribution match). As a result, a statistical model
that is constructed using a Poisson distribution has the potential
to have generated the observed data. Adhering to this principal results
in generative statistical models that capture important characteristics
of the process (e.g., predicted counts that are always $\geq0$),
which can be important for interpretation and model checking (\citealt{conn2018guide,gelfand2018bayesian}).
When constructing a statistical model for distance sampling data probability
distributions should match the following characteristics of the data:
1) the number and locations of individuals are random variables; and
2) the location of individuals exists in continuous geographic space.
In what follows, we use the term \textquotedblleft continuous geographic
space\textquotedblright{} to describe spatial areas that are defined
as polygons and contain an infinite number of possible locations (points)
within the boundary of each polygon. 

Researchers have developed model-based approaches for the spatial
analysis of distance sampling data (e.g., \citealt{miller2013spatial,buckland2016model}),
but existing approaches do not account for location uncertainty except
in the case where the distribution of individuals is spatially uniform
(e.g., \citealt{borchers2015unifying}). Our approach builds upon
one of the most common model-based methods for the spatial analysis
of distance sampling data that uses an inhomogeneous Poisson point
process (IPPP) distribution, which allows for heterogeneity in the
spatial distribution of individuals. In what follows, we review previously
developed modeling approaches based on the IPPP distribution and then
extend this model to account for location uncertainty.

The IPPP distribution describes the random number and locations of
individuals within a continuous geographic space. The IPPP is constructed
by assuming the spatial distribution of individuals is explained by
an intensity function, $\lambda(\mathbf{s})$, where $\mathbf{s}$
is a vector that contains the coordinates of a single location contained
within the study area $\mathcal{S}$. The intensity function is commonly
specified as 
\begin{equation}
\text{log}(\lambda(\mathbf{s}))=\beta_{0}+\boldsymbol{\mathbf{x}}(\mathbf{s})^{'}\boldsymbol{\beta}\:,
\end{equation}
where $\beta_{0}$ is the intercept, $\boldsymbol{\mathbf{x}}(\mathbf{s})$
is a $p\times1$ vector that contains predictor variables at location
$\mathbf{s}$, and $\boldsymbol{\beta}\equiv(\beta_{1},...,\beta_{p})^{'}$
is a vector of regression coefficients. Estimating the regression
coefficients from distance sampling data enables inference regarding
species-habitat relationships. 

An important property of the IPPP distribution is that estimates of
abundance can be obtained for any geographic region that is contained
within the study area. More precisely, for any sub-region $\mathcal{A}$
contained within the study area $\mathcal{S}$, an estimate of abundance
is 
\begin{equation}
\bar{\lambda}=\intop_{\mathcal{A}}\lambda(\mathbf{s})d\mathbf{s}\:,
\end{equation}
which is referred to as the integrated intensity function. Clearly,
accurate estimates of abundance requires reliable estimation of the
intensity function ($\lambda(\mathbf{s})$) which depends on the intercept
($\beta_{0}$), regression coefficients ($\boldsymbol{\beta}$) and
predictor variables ($\boldsymbol{\mathbf{x}}(\mathbf{s})$).

As with traditional distance sampling methods, the IPPP can incorporate
a detection function, which we denote $q(\mathbf{s})$, where $q(\cdot)$
is the usual detection function (e.g., half-normal function) that
depends on the location $\mathbf{s}$ by way of the distance between
an individual and the point or line transect. Employing the notation
from above, the probability distribution function for the IPPP is
\begin{equation}
[\mathbf{z}_{1},\mathbf{z}_{2},...,\mathbf{z}_{n},n|\lambda(\mathbf{s}),q(\mathbf{s})]=e^{\mathbf{-\intop_{\mathcal{S}}\lambda(\mathbf{s})}q(\mathbf{s)}d\mathbf{s}}\prod_{i=1}^{n}\lambda(\mathbf{z}_{i})q(\mathbf{z}_{i})\:,
\end{equation}
where $\mathbf{z}_{1},\mathbf{z}_{2},...,\mathbf{z}_{n}$ are the
coordinate vectors (i.e., exact locations) of the $n$ detected individuals
(\citealt{johnson2010model}). The product of $\lambda(\mathbf{s})$
and $q(\mathbf{s})$ is referred to as the thinned intensity function
(\citealt{cressie1993statistics}, p. 689). The bracket notation $[\cdot]$,
used on the left hand side of Eq. 3, represents a probability distribution.
Using bracket notation, $[y,z]$ is a joint distribution where $y$
and $z$ are the random variables, $[y|z]$ is a conditional distribution
where $y$ is the random variable given $z$. The marginal distribution
of $y$ can be obtained by ``integrating out'' $z$ (i.e., $[y]=\int[y,z]dy$).

When expressed as a likelihood function, Eq. 3 can be used to estimate
parameters associated with the intensity and detection functions.
For example, using Eq. 3 as a likelihood function facilitates estimation
of the regression coefficients, $\boldsymbol{\beta}$, from Eq. 1
and enables inference regarding species-habitat relationships. Evaluation
of the likelihood function, however, requires the exact locations
of all $n$ detected individuals.

\vspace{0.66cm}

\begin{spacing}{1.9}

\subsection*{ACCOUNTING FOR UNKNOWN LOCATION}
\end{spacing}

\noindent Parameter estimation and statistical inference using the
IPPP distribution requires the exact location of each detected individual
because the likelihood function from Eq. 3 assumes that $\mathbf{z}_{i}$
is recorded. When collecting distance sampling data, the exact location
of each detected individual is usually not recorded which generates
location uncertainty. Below, we extend the IPPP distribution so that
the model can be implemented when the location of individuals is uncertain.
Our extension could be viewed as a special case of the unified approach
of \citet{borchers2015unifying}, however, the authors did not present
how models that involve nonuniform distribution of plants and animals,
such as the IPPP, might be implemented. In what follows, we show in
detail how to implement such models.

Using distance sampling data collected from a line transect, an individual's
exact location is an unknown point that lies on one of two lines parallel
to the transect at a perpendicular distance that is equal to the recorded
distance. Similarly, for point transects, the location of an individual
is an unknown point on the perimeter of a circle centered on the transect
with a radius that is equal to the recorded distance. If the exact
location is unknown but lies on a line or perimeter of a circle, a
model that is based on the IPPP distribution and accounts for location
uncertainty in
\begin{equation}
[\mathbf{y}_{1},\mathbf{y}_{2},...,\mathbf{y}_{n},n|\lambda(\mathbf{s}),q(\mathbf{s})]=e^{\mathbf{-\intop_{\mathcal{S}}\lambda(\mathbf{s})}q(\mathbf{s)}d\mathbf{s}}\prod_{i=1}^{n}{\displaystyle |\mathcal{L}_{i}|}^{-1}\intop_{\mathcal{L}_{i}}\lambda(\mathbf{z}_{i})q(\mathbf{z}_{i})d\mathbf{z}_{i}\:
\end{equation}
where the modification involves replacing the product of $\lambda(\mathbf{z}_{i})$
and $q(\mathbf{z}_{i})$ in Eq. 3 with an integral. In Eq. 4, the
random variable is the coordinate vectors $\mathbf{y}_{1},\mathbf{y}_{2},...,\mathbf{y}_{n}$
and number of detected individuals $n$. As in Eq. 3, $\mathbf{z}_{i}$
is the exact coordinate of the $i^{\text{th}}$ individual, which
is integrated out of the joint distribution to obtain Eq. 4. Knowing
the distance and transect of detection determines the limits of integration
in Eq. 4, where $\mathcal{L}_{i}$ is the parallel lines (circle perimeter)
and $|\mathcal{L}_{i}|$ is the length of the lines (or circle perimeter).
Conceptually $\mathbf{y}_{i}$ can be thought of as the coordinate
where the $i^{\text{th}}$ individual was ``recorded'', which is
different than the true location of the individual $\mathbf{z}_{i}$
because the ``recorded'' location is a uniformly distributed point
on $\mathcal{L}_{i}$ (see Appendix S1 for more detail).\vspace{0.66cm}

\subsection*{ACCOUNTING FOR MEASUREMENT ERROR IN DISTANCES}

In many cases the distance between the transect and the detected individuals
may be recorded with error, which introduces another source of location
uncertainty. For example, in the data illustration that follows, the
distances recorded for individual birds close to the transect line
were almost certainly recorded with greater accuracy than those detected
further from the line. To account for error in distances, we construct
a hierarchical model where the observed random variable, $\mathbf{y}_{i}$,
is the ``recorded'' location which depend on the exact locations
$\mathbf{z}_{i}$ (see Appendix S1 for more detail). A hierarchical
model that accounts for both types of location uncertainty is
\begin{equation}
[\mathbf{y}_{1},\mathbf{y}_{2},...,\mathbf{y}_{n},n|\boldsymbol{\theta},\lambda(\mathbf{s}),q(\mathbf{s})]=e^{\mathbf{-\intop_{\mathcal{S}}\lambda(\mathbf{s})}q(\mathbf{s)}d\mathbf{s}}\prod_{i=1}^{n}{\displaystyle |\mathcal{L}_{i}|}^{-1}\intop_{\mathcal{S}}[d(\mathbf{y}_{i},t_{i})|d(\mathbf{z}_{i},t_{i}),\boldsymbol{\theta}]\lambda(\mathbf{z}_{i})q(\mathbf{z}_{i})d\mathbf{z}_{i}\:.
\end{equation}
In equation above, $\boldsymbol{\theta}\equiv(\theta_{1},\theta_{2},...\theta_{m})^{'}$
is a vector of unknown parameters of $[d(\mathbf{y}_{i},t_{i})|d(\mathbf{z}_{i},t_{i}),\boldsymbol{\theta}]$,
which is a probability distribution that describes the recorded distances
$d(\mathbf{y}_{i},t_{i})$ given the true distances $d(\mathbf{z}_{i},t_{i})$.
The function $d(\cdot,\cdot)$ returns the perpendicular distance
between a coordinate vector and the transect, $t_{i}$, where the
$i^{\text{th}}$ individual was detected. We refrain from specifying
the functional form of $[d(\mathbf{y}_{i},t_{i})|d(\mathbf{z}_{i},t_{i}),\boldsymbol{\theta}]$
because this portion of our model is general and any appropriate distribution
can be chosen as shown in the data illustration.

\subsection*{MODEL IMPLEMENTATION}

For all models, estimating the parameters associated with the intensity
($\mathbf{\lambda(\mathbf{s})}$) and detection ($q(\mathbf{s})$)
functions requires evaluation of the the integrals in the likelihood.
In nearly all situation, solving the integrals will require numerical
integration using approximations such as Monte Carlo or numerical
quadrature (\citealt{BB2L}). Using a numerical quadrature involves
the approximation
\begin{equation}
\intop_{\mathcal{A}}f(\mathbf{s)}d\mathbf{s}\approx|\mathcal{A}|{\displaystyle \sum_{q=1}^{Q}}f(\mathbf{s}_{q})\:,
\end{equation}
where $f(\mathbf{s})$ is an unspecified function, $\mathcal{A}$
is an arbitrary region (or line) with area (or length) $|\mathcal{A}|$
and $Q$ is the number of (equally spaced) points partitioning the
polygon (or line). The function $f(\mathbf{s})$ is specified based
on the integral. For example, Eqs. 3\textendash 5 contains the integral
$\mathbf{\intop_{\mathcal{S}}\lambda(\mathbf{s})}q(\mathbf{s)}d\mathbf{s}$,
which could be approximated by defining $f(\mathbf{s})\equiv\lambda(\mathbf{s})q(\mathbf{s}).$

Accounting for location uncertainty requires a hierarchical model
because the probability distribution in Eqs. 4 and 5 were constructed
by conditioning on the exact locations which are random variables
that follow an IPPP distribution (see Appendix S1). Regardless of
the inferential paradigm, the models are challenging to fit to distance
sampling data because each detected individual has a latent (unobserved)
true coordinate vector, which results in $2n$ additional parameters.
For example, the true coordinate vectors could be estimated by fully
specifying a Bayesian model and obtaining samples from the marginal
posterior using a Markov chain Monte Carlo (MCMC) algorithm. This
approach, however, requires sampling from a high-dimensional posterior
distribution. Such an approach is challenging because the MCMC algorithm
can be difficult to tune and require multiple evaluations of the likelihood
function, involving a quadrature approximation. 

Estimates of the true coordinate vectors are of little interest because
most studies use distance sampling data to obtain predictions densities
and infer species-habitat relationships. As a result, the true coordinate
vectors can be treated as ``nuisance parameters'' and removed by
integrating the joint likelihood as we did in Eqs. 4\textendash 5
(\citealt{borchers2015unifying}). The integrated likelihood has $2n$
fewer parameters and the remaining parameters can be estimated using
maximum likelihood estimation or by sampling from the posterior distribution
using techniques such as MCMC (see Appendix S1).

In both the simulation experiment and data example that follow, we
estimate all parameters by maximizing the appropriate likelihood using
the Nelder-Mead algorithm in R (\citealt{nelder1965simplex,TeamR};
see Appendix S2 and S3). For all model parameters, we obtain approximate
variances by inverting the Hessian matrix and construct Wald-type
confidence intervals (CIs; \citealt{pawitan2001all}). To obtain CI
for derived quantities, we use percentiles of the empirical distribution
obtained from a bootstrapping approach outlined by \citet{lele2010estimability}
based on the results of \citet{harris1989predictive}.\vspace{0.66cm}

\begin{spacing}{1.9}

\subsection*{SIMULATION EXPERIMENT}
\end{spacing}

\begin{doublespace}
\noindent We conducted a simulation experiment to evaluate the influence
of location uncertainty on model-based distance sampling methods and
to test the efficacy of our new approach. We expect that standard
approaches will result in biased parameter estimates which may obscure
species-habitat relationships or, in the worst case, result in misleading
conclusions. Conversely, we expect that our proposed method that accounts
for location uncertainty will result in unbiased parameter estimates
(in the asymptotic sense). We expect that accounting for location
uncertainty will result in parameter estimates that are more variable
and have estimates of uncertainty that are appropriately inflated
(e.g., CIs will be wider) when compared to estimates obtained from
exact locations.
\end{doublespace}

We simulated the exact location of individuals from an IPPP distribution
on the unit square with a single predictor variable ($x(\mathbf{s})$)
and specified the intensity function as $\text{log}(\lambda(\mathbf{s}))=\beta_{0}+\beta_{1}$$x(\mathbf{s})$
with $\beta_{0}=9$ and $\beta_{1}=1$ (Fig. 1). We evaluated two
scenarios by varying the location of 16 point transects. In the first
scenario, we placed point transects in poorer quality ``habitat''
(i.e., at lower values of $x(\mathbf{s})$; Fig. 1a) whereas in the
second scenario we randomly placed the transects but restricted transect
placement so that detection of the same individual from multiple transect
was not possible (Fig 1b). We designed the first and second scenarios
to evaluate how location uncertainty influences parameter estimates
when transects are placed based on convenience and under a randomized
design respectively. We simulated the detection of each individual
using a Bernoulli distribution by calculating the probability of detection
for each individual using a truncated half-normal detection function
specified as $p_{i}=e^{-\left(\frac{d_{i}}{0.025}\right)^{2}}I(d_{i}<0.06),$
where $p_{i}$ is the probability of detection for the $i^{\text{th}}$
individual that occurs at a distance $d_{i}$ from the point transect
(Fig. 1). 

We simulated 250 data sets for each scenario and fit four models to
each data set. For the first model, we used the exact locations of
the individuals and fit Eq. 3 to the simulated data. This is the ideal
situation because the generating process used to simulate the data
matches the generating process specified by the statistical model.
Thus, we expected unbiased parameter estimates with the narrowest
CIs under model 1. For the second and third models, the ``available''
data included only the distances from the point transects to the locations
of the detected individuals. Because the exact locations of the individuals
were unavailable, we fit Eq. 3 using two different surrogate predictors
which included: 1) the value of $x(\mathbf{s})$ at the point transect
where the individual was detected (model 2); and 2) the average of
$x(\mathbf{s})$ within a distance of $0.06$ of the transect where
the individual was detected (model 3). The distance of $0.06$ distance
was chosen to correspond to the value used to truncate the half-normal
detection function, which would be unknown in practice. Finally, our
fourth model uses the same data as the second and third model, but
instead accounts for location uncertainty using Eq. 4. 

For each of the two scenarios and four models, we assessed reliability
of the inferred species-habitat relationship by comparing the true
values of $\beta_{1}$ to the estimated value and the coverage probability
of the 95\% CIs. We assessed efficiency by calculating the average
length of the 95\% CI for $\hat{\beta}_{1}$ that was obtained from
the model that accounts for location uncertainty (i.e., model 4) divided
by the average length of the 95\% CI for $\hat{\beta}_{1}$ obtained
from fitting the IPPP distribution using data with exact locations
(i.e., model 1). In Appendix S2, we provide a tutorial with R code
to implement the simulation and reproduce Fig. 1 and table 1 .\vspace{0.66cm}

\begin{spacing}{1.9}

\subsection*{FIELD-COLLECTED AVIAN DATA}
\end{spacing}

\noindent Distance sampling data from 137 bird species were collected
over a 29 year period from 1981 to 2009 as part of the Long Term Ecological
Research Program at the Konza Prairie Biological Station (KPBS; Fig.
2). The KPBS is a tallgrass prairie site located in northeastern Kansas,
USA and is experimentally managed under varied grazing and fire regimes
(\citealt{knapp1998grassland}). We used data from a single species,
Dickcissels (\textit{Spiza americana}), which are the most common
grassland-breeding species at KPBS. Both male and female Dickcissels
perch conspicuously from the tops of vegetation and males vocalize
frequently (\citealt{temple2002}). For our analysis, we used observations
of Dickcissels collected by a single observer over the period of May
27, 1981 to June 26, 1981. This resulted in 106 individuals detected
on 11 of the 14 transects at perpendicular distances ranging from
$0\,\text{m}$ to $61\,\text{m}$. A full description of the data
is provided in \citet{zimmerman1993birds} and \citet{KONZADATA}.

We illustrate our method using elevation as a predictor variable (Fig.
2). Elevation within the KPBS is available from a digital elevation
model that has cell resolution of $2\,\text{m}\times2\,\text{m}$
(\citealt{KONZAElev}). Based on previous research, we expect the
abundance of Dickcissels should be greater at higher elevations when
compared to lower elevations (\citealt{zimmerman1993birds}). Given
our distance sampling data, it is not possible to reconstruct the
exact location of each detected individual, therefore, we are unable
to obtain the elevation at the location of each detected Dickcissel.
We implement four models that included: a) the standard distance sampling
model (Eq. 3) using a surrogate predictor which was the average elevation
along the transect where the individual was detected (model a); b)
a model that accounts for location uncertainty assuming that distances
are recorded without error (i.e., Eq. 4; model b); c) a model that
accounts for location uncertainty and distance mismeasurement that
follows a truncated normal distribution (model c); and d) a model
that accounts for location uncertainty distance mismeasurement that
follows a Laplace distribution. For models c and d, which accounted
for error in the recorded distances, we assumed that variance of the
normal and Laplace distributions was zero on the transect line, but
increased linearly at an unknown rate as the distance between the
individual bird and transect increased (see Fig 3c. for an example).
We truncated the normal and Laplace distributions at distances below
$0\,\text{m}$ and above $150\,\text{m}$ to increase computational
efficiency of the quadrature approximation. Because the maximum recorded
distance in our data set was $61\,\text{m}$, this truncation has
negligible influence on our results. Depending on the specifics of
the study design, it is easy to incorporate different specifications
such as a constant variance model or alternative probability distributions
(e.g., a Poisson distribution for distances that are rounded to the
nearest meter). In Appendix S3, we include additional details associated
with the field-collected avian data analysis along with a tutorial
and R code to implement the models and reproduce Table 2 and Figs.
2 and 3.\vspace{0.66cm}

\section*{Results}

\subsection*{SIMULATION EXPERIMENT}

When the exact location of each detected individual was available,
the standard IPPP model in Eq. 3 (model 1) performed as expected in
that estimates of $\beta_{1}$ appeared to be unbiased and 95\% CI
covered the true value with probabilities between 0.93\textendash 0.96
(Fig 1; Table 1). In contrast, when location uncertainty was not accounted
for (models 2 and 3), the estimated regression coefficients were biased
(Fig. 1) and coverage probabilities of the 95\% CIs were $\leq0.38$
(Table 1). When the transect locations were placed based on convenience
and the surrogate predictor variable was obtained from the center
of the transect (model 2), the bias was particularly large and resulted
in negative estimates of regression coefficients for most data sets
even though the true value was $\beta_{1}=1$ (Fig. 1a). Our method
(model 4), which accounted for location uncertainty, yielded apparently
unbiased estimates of $\text{\ensuremath{\beta_{1}}}$ for both scenarios
and produced coverage probabilities between $0.95$\textendash $0.98$
(Fig. 1; Table 1). The 95\% CI were $1.49$\textendash $1.50$ times
longer when the exact location was unknown (Table 1). These results
demonstrate that our proposed method efficiently accounted for location
uncertainty and resulted in parameter estimates that are about 50\%
less precise than parameter estimates obtained when the exact location
is known. This shows that the loss of information resulting from location
uncertainty could be ameliorated by collecting more data.\vspace{0.66cm}

\begin{spacing}{1.9}

\subsection*{FIELD-COLLECTED AVIAN DATA}
\end{spacing}

All three models that accounted for location uncertainty (models b\textendash d)
had similar Akaike information criterion scores that were $>811$
points lower than model a, which used Eq. 3 and the average elevation
along the transect as the predictor (Table 1). This indicates that
accounting for location uncertainty improved the fit of the model
to the data. Predicted Dickcissels abundance at higher elevations
was greater when location uncertainty was accounted for in models
b\textendash d (Table 2; Figure 3a). This difference in predictions
was caused by the regression coefficient estimates, which were 27\%
larger for the three models that accounted for location uncertainty
(models b\textendash d) when compared to the model that used the surrogate
predictor (model a; Table 2).\textcolor{black}{{} The model comparison
of the estimated relationship between elevation and abundance, distance
and the probability of detection, and the true distance and recorded
distance are visualized in Fig. 3.}

\section*{Discussion}

This study demonstrates that location uncertainty, when unaccounted
for, can result in biased parameter estimates and unreliable inference
regarding species-habitat relationships. Within a broader context,
location uncertainty manifests as an ecological fallacy namely that
the inferred relationship from aggregated data could differ when compared
to individual-level data (\citealp{ecologicalfallacy,cressie2011statistics},
p. 197). Point process models using exact locations of individuals
targets inference about the habitat and environmental preferences
of individuals whereas ignoring location uncertainty and using transect-wide
surrogate predictors results only in inference about how abundance
varies among the transects. Individual-level inference is invariant
to changes in the spatial scale of the analysis, whereas transect-level
inference is scale specific. Our study demonstrates that spatial model-based
approaches for traditional distance sampling data can provide reliable
individual-level inference, even when the exact locations of individuals
are unknown. This is a significant advancement because our approach
enables individual-level inference, but does not require auxiliary
data that may difficult or impossible to obtain for historical data
sets (e.g., \citealt{borchers2015unifying}).

Our approach also offers insight into best practices for future distance
sampling study design when recording the exact location of individuals
may be difficult or infeasible. Our simulation results suggest that,
given a desired statistical power or level of precisions for parameter
estimates, there is a tradeoff between sample size (i.e., the number
of detected individuals) and location accuracy. The deleterious effects
of collecting distance sampling data without recording the exact location
can be remediated by simply collecting more data and selecting an
appropriate model. For example, in both scenarios of our simulation
experiment, the same precision of parameter estimates can be achieved
by either detecting $n$ individuals and recording their exact locations,
or by detecting $\approx1.5n$ individuals and recording only their
distances. Although the sample size calculations from our simulation
results are not generalizable to future data collection efforts, study-specific
power analyses could be conducted to determine the tradeoff between
the two data collection approaches. \vspace{0.66cm}

\begin{spacing}{1.9}

\subsection*{PRACTICAL GUIDANCE FOR DATA ANALYSIS}
\end{spacing}

\noindent Location uncertainty is ubiquitous in all sources of spatially
referenced data because it is impossible to measure and record the
location with infinite precision. Despite the presence of location
uncertainty in all sources of data, accounting for it may be time
consuming because the models are tailored to the specifics of the
study and usually must be fit using ``custom built'' computer code
(\citealt{BB2L}). For some data sets accounting for location uncertainty
will be required and in other data sets it may not be possible or
beneficial. Prior to fitting models to distance sampling data, we
urge researchers consider the seven questions below to determine possible
impact of location uncertainty on study outcomes.
\begin{enumerate}
\item \textit{Does the predictor variable exhibit fine-scale spatial variability?}
If so, the predictor variable is likely to change when moving a short
distance from one location to another location. In this case, the
surrogate predictor variable (e.g., elevation at the transect centroid)
will likely differ from the predictor variable at the location of
the individuals. Whenever the surrogate predictor variable included
in a model is different from the value of the predictor at the exact
locations, there is the potential for bias. The larger the difference
between the value of the surrogate variable and the value at the exact
location, the more important it will be to account for location uncertainty.
\item \textit{Are the placement of the transects related to the spatial
structure of the predictor variable?} For example, the transects may
be placed along roads within larger areas of homogeneous habitat.
In this case, a surrogate predictor such as the percentage of grassland
within $100\,\text{m}$, may be strongly influenced by the location
of the transects and creates a potential for bias to occur. Accounting
for location uncertainty may be needed.
\item \textit{Are the spatial scales of the predictor variables known?}
In many cases environmental characteristics within an area surrounding
the location of the individuals is used to determine the influence
of landscape-level processes (e.g., percentage of grassland within
$100\,\text{m}$). These approaches use a buffer or kernel that is
typically centered at the exact location of the individual. The predictor
variable is calculated by integrating the kernel and point-level predictor
variable over the study area (e.g., \citealt{heaton2011kernel,heaton2011spatial,heaton2012kernel,bradter2013identifying,chandler2016estimating,miguet2017quantify,stuber2017bayesian}).
Because the buffer or kernel is centered at the exact location of
the individuals, accounting for location uncertainty is likely to
influence the inference. 
\item \textit{Is the spatial resolution of the predictor variables too course?}
In some situation the spatial resolution of the predictor variable
will only be available at a course grain. For example, WorldClim provides
a set of climate variables that are predictions available on a $1\times1\,\text{km}$
grid (\citealt{hijmans2005very}). The transect from our data example
are all $\leq1589\,\text{m}$ in length. For a given transect, most
detected individuals would be assigned the same value of the predictor
variables from WorldClim because most of the individual birds occur
within a single $1\times1\,\text{km}$ grid cell. If the goal of the
study is to relate abundance to climatic variables using WorldClim,
then in such a case, the researcher would experience minimal or no
gain from accounting for location uncertainty.
\item \textit{Are spatial data for the predictor variables available over
the entire study area?} In some situations the predictor variables
will not be available at every location within the study area. For
example, many studies that collect distance sampling data on animals
also collect detailed information on vegetation at a feasible number
of locations within the study area. It is tempting to use vegetation
measurements as predictor variables that are collected at the location
that is thought to be closest to the individual animal. This approach
presents two challenges: 1) if the vegetation at the location of the
individual is different than the location where the measurements were
taken, the predictor variables may result in biased coefficient estimates;
and 2) fitting point process models to data requires a continuous
surface of the predictor variable over the entire study area. In this
situation, we recommend first building a statistical model that can
predict the vegetation measurements as a continuous surface over the
study area. This is equivalent to building a custom high-resolution
``raster layer'' using the the vegetation data. Developing auxiliary
models for predictor variables that are measured in the field is a
common technique used in spatial statistics to ameliorate the problem
of misaligned data (\citealt{gotway2002combining,gelfand2010misaligned};
\citeauthor{Pacifici2019} \textit{in press}). Once the predictor
variable is available as a continuous surface or high resolution raster
layer, then accounting for location uncertainty is likely to be beneficial. 
\item \textit{Is the location uncertain for only a portion of the observations?}
There may be situations were only a portion of the observations have
uncertain locations (e.g., \citealt{hefley2014correction,hefley2017bias}).
If the number of observations with uncertain locations is small (e.g.,
$<5\%$ of the observation), these could be removed from the data
set and perhaps cause only minor changes in inference. If the number
of observations with uncertain locations is larger, then we recommend
constructing models that integrate both sources of data (e.g., by
combining portions of the likelihood in Eq. 3 and Eq. 4). Similar
approaches could be applied to situations where different sources
of data result in the magnitude of the location uncertainty being
variable. For example, the mismeasurement of distances in our historic
Dickcissel distance sample data are likely to be larger than more
recent surveys in the same data set because researchers adopted the
use of laser rangefinders.
\item \textit{Do the predictor variables contain errors?} In some situations,
the predictor variables contain errors. For example, researchers may
include modeled climatic predictor variables, but the predicted climate
at any given location is different than the true conditions. In this
situation, the error in the predictors may mask or exacerbate the
effect of location uncertainty. This problem is well-studied in the
statistics literature where it is known as ``errors-in-variables''
(\citealt{carroll2006measurement}). In addition to location uncertainty,
errors-in-variables can be accounted for by using a hierarchical modeling
framework (e.g. \citealt{foster2012uncertainty,stoklosa2015climate,hefleyHSDM}).\vspace{0.66cm}
\end{enumerate}
\begin{spacing}{1.9}

\subsection*{CONCLUSION}
\end{spacing}

\noindent Ecological data are messy in ways that result in many potential
biases. For example, failing to account for detection may result in
biased estimates of abundance. Ecologists have focused intensively
on some sources of bias (e.g., detection) while paying little attention
to other sources of bias such as location uncertainty. Thus, there
is a shortage of tools available for less studied sources of bias,
leading researchers to rely on ad hoc and untested procedures. We
recommend avoiding untested procedures because location uncertainty
can create complex biases that are difficult to anticipate and understand
(e.g., \citealp{montgomery2011implications,hefley2014correction,brost2015animal,mitchell2016sensitivity,hefley2017bias,gerber2018accounting,walker2018bias};
\citeauthor{walker2019bias} \textit{under revision}). Model-based
approaches have provided a wealth of tools to address biases in many
types of ecological data. In situation where location uncertainty
is a concern, model-based approaches like those proposed in our study
will enable researchers to make reliable ecological inference and
accurate predictions.

\noindent \vspace{0.66cm}

\section*{Acknowledgements}

We thank all individuals, including Elmer Finck, John Zimmerman, and
Brett Sandercock, who contributed to the the distance sampling data
used in our study. The material in this study is based upon research
supported by the National Science Foundation (DEB-1754491 and DEB-1440484
under the LTER program).\vspace{0.66cm}

\section*{Data accessibility}

The field-collected avian data is available from \citet{KONZADATA}.
Additional files required to reproduce the results of this study (e.g.
shapefiles of transects) are archived in the Dryad Digital Repository
(\citealt{hefley2019dryaddata}).

\renewcommand{\refname}{\vspace{0.1cm}\hspace{-0cm}\selectfont\large References}\setlength{\bibsep}{0pt}

\bibliographystyle{apa}
\bibliography{refs}
\vspace{-0.5cm}

\section*{Supporting Information}

\begin{singlespace}
\noindent Additional Supporting Information may be found in the online
version of this article.
\end{singlespace}

\noindent \textbf{Appendix S1}

\noindent In-depth description and explanation of point process models
for distance sampling data.

\noindent \textbf{Appendix S2}

\noindent Tutorial with R code to reproduce the simulation experiment,
table 1, and figure 1.

\noindent \textbf{Appendix S3 }

\noindent Tutorial and R code to reproduce the data example, table
2, and figures 2 and 3.\pagebreak{}

\noindent \textbf{Fig. 1.} Our simulation experiment used distance
sampling data collected from transects placed by convenience (panel
a) and using a randomized design (panel b). The predictor variable,
$x(\mathbf{s})$, represents ``habitat'' where black shading is
preferred habitat and the maximum values of $x(\mathbf{s})$ and white
is avoided habitat the minimum values of $x(\mathbf{s})$. Individuals
(red \mycirc[red]) were sampled by placing 16 point transects (black
$\diamond$), and individuals within the larger black circles were
available for detection. Detected individuals (blue \textcolor{blue}{x})
were the data used for our simulation experiment. Data was simulated
250 times for each sample design and for each data set we fit four
model-based approaches which included the following: 1) Eq. 3 with
the exact location of individuals known (Exact; model 1); 2) Eq. 3
with the exact location unknown and a surrogate predictor that was
the value of $x(\mathbf{s})$ at the center of the transect (Transect
center; model 2); 3) Eq. 3 with a surrogate predictor that was the
average value of $x(\mathbf{s})$ within the black circle (Transect
average; model 3); and 4) Eq. 4 that does not require the exact location
of the individuals to be known (Corrected; model 4). The violin plots
on the right show the estimated regression coefficients $\beta_{1}$
obtained from the 250 simulated data sets. The true value of $\beta_{1}=1$
is identified by the black dashed line.\bigskip{}

\noindent \textbf{Fig. 2.} Line transects used to collect distance
sampling data at Konza Prairie Biological Station. The transects were
placed to sample watershed-level experimental treatments.\\
\bigskip{}

\noindent \textbf{Fig. 3.} Model-based estimates (dashed lines) and
95\% CIs (colored shading) showing the relationship between elevation
and expected abundance of Dickcissel (panel a), distance and the probability
of detection (panel b), and the true distance and recorded distance
(panel c). Estimates were obtained from a model that does not account
for location uncertainty and uses the average elevation (red; model
a in methods) and a model that accounts for location error and assumes
normally distributed distance errors (blue; model c). The black tickmarks
on the y-axis of panel c are the recorded distances between the line
transects and individual Dickcissels.

\pagebreak\begin{landscape}
\begin{figure}[H]
\begin{centering}
\includegraphics{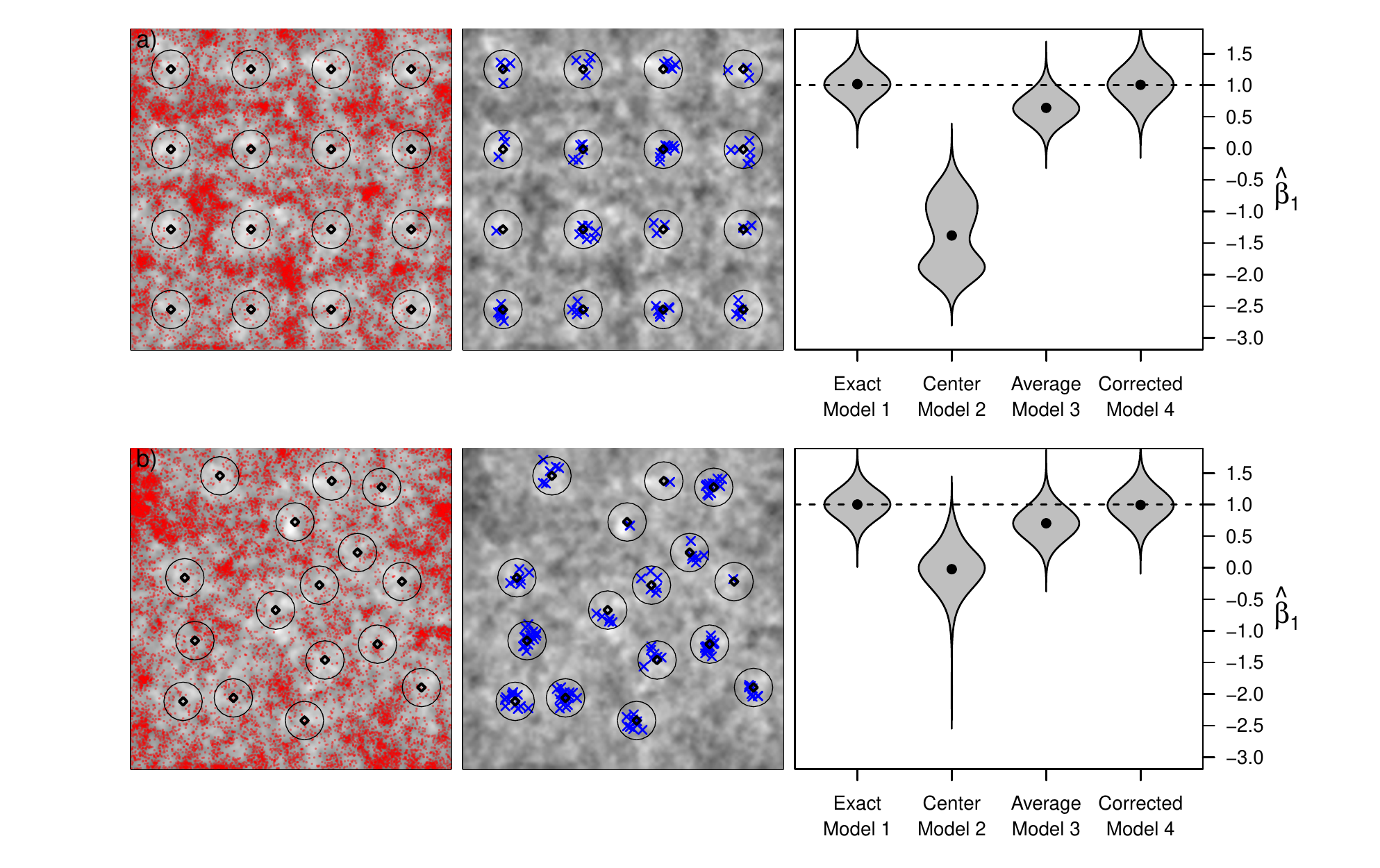}
\par\end{centering}
\caption{}
\end{figure}
\begin{figure}[H]
\begin{centering}
\includegraphics{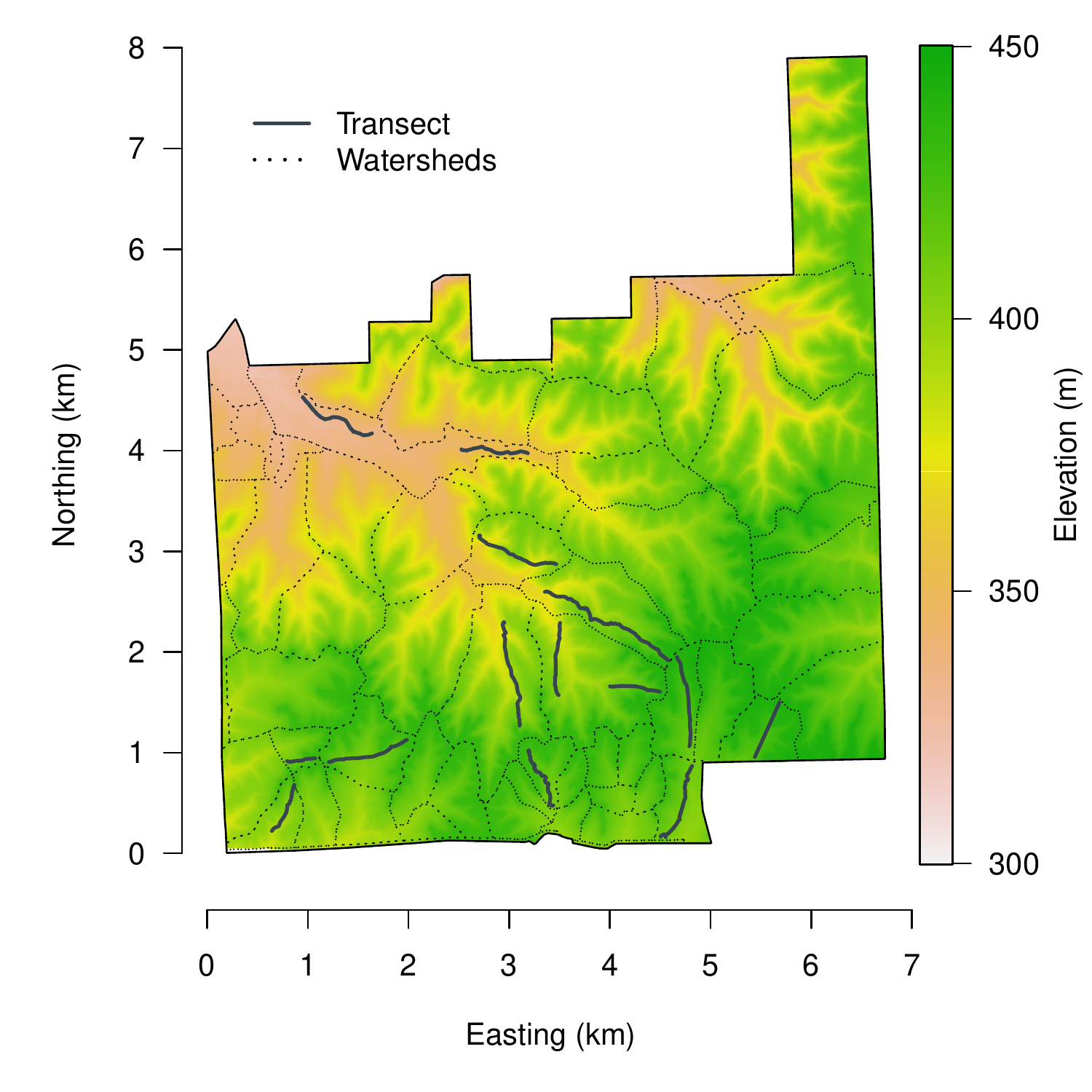}
\par\end{centering}
\caption{}
\end{figure}
\begin{figure}[H]
\begin{centering}
\includegraphics[scale=1.3]{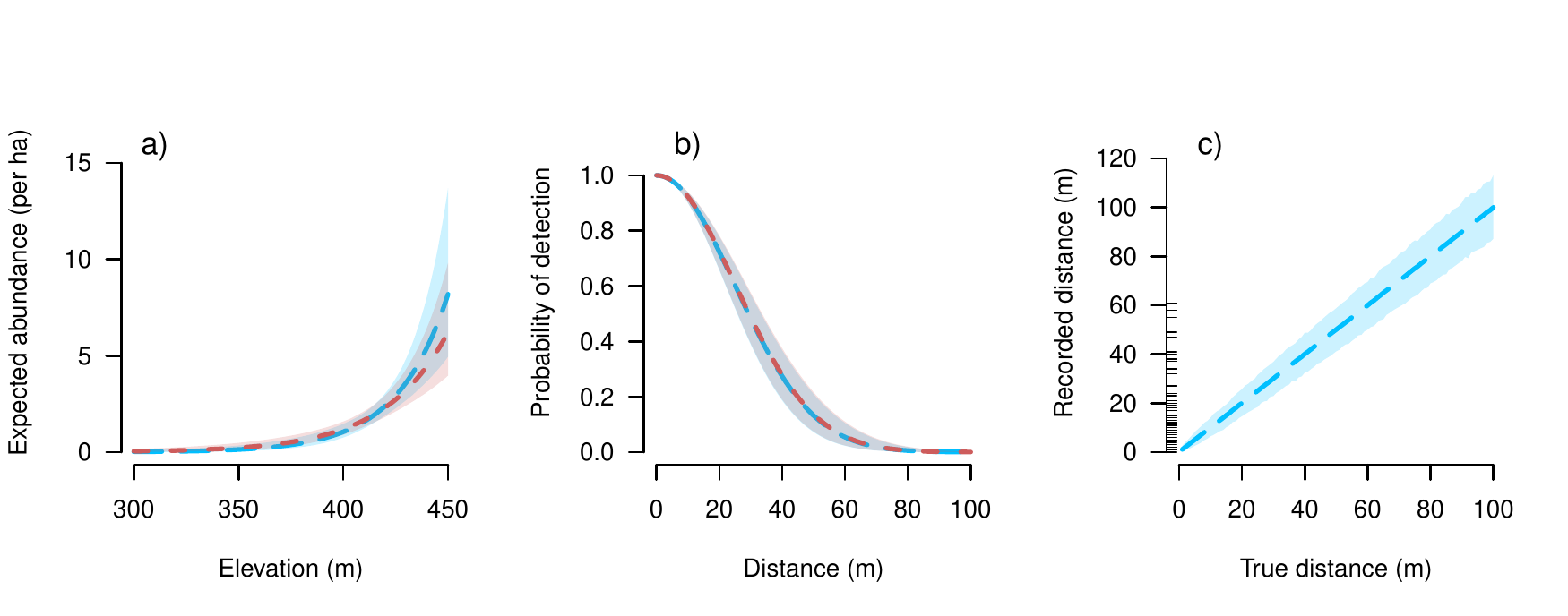}
\par\end{centering}
\caption{}
\end{figure}
\end{landscape}

\pagebreak{}

\begin{landscape}

\noindent \textbf{Table 1}. Results of our simulation experiment that
evaluated the influence of location uncertainty. For each scenario,
we simulated 250 data sets and varied the sampling design by placing
transect based on convenience or randomly (see Fig. 1 for illustration).
We report the average number of detected individuals ($\bar{n}$)
for each scenario. For each simulated data set, we fit four models
which included: 1) Eq. 3 with the exact location of individuals known
(model 1); 2) Eq. 3 using a surrogate predictor that was obtained
from the center of the transect (model 2); 3) Eq. 3 with a surrogate
predictor that was the transect average (model 3); and 4) Eq. 4 that
does not require the exact location of the individuals to be known
(model 4). We report the estimated coverage probability (CP) for the
95\% confidence interval (CI) for the regression coefficient ($\beta_{1}$
in Eq. 1) estimated using each model. We calculated efficiency by
dividing the average length of the 95\% CI of $\hat{\beta}_{1}$ from
the location uncertainty corrected model (model 4) by average length
of the 95\% CI of $\hat{\beta}_{1}$ estimated from Eq. 3 with the
exact locations known (model 1). 

\begin{center}
\begin{tabular*}{1.4\textwidth}{@{\extracolsep{\fill}}>{\raggedright}p{0.09\textwidth}>{\raggedright}p{0.17\textwidth}>{\raggedright}p{0.07\textwidth}>{\raggedright}p{0.18\textwidth}>{\raggedright}p{0.18\textwidth}>{\raggedright}p{0.18\textwidth}>{\raggedright}p{0.18\textwidth}>{\raggedright}p{0.12\textwidth}}
\hline 
Scenario & Transect & $\bar{n}$ & CP (model 1) & CP (model 2) & CP (model 3) & CP (model 4) & Efficiency\tabularnewline
\hline 
1 & Convenience & 94 & 0.96 & 0.00 & 0.26 & 0.95 & 1.50\tabularnewline
2 & Random & 124 & 0.93 & 0.00 & 0.38 & 0.98 & 1.49\tabularnewline
\hline 
\end{tabular*}
\par\end{center}

\end{landscape}

\pagebreak{}

\begin{landscape}

\noindent \textbf{Table 2}. Parameter estimates and 95\% confidence
intervals (in parentheses) associated with the intercept ($\beta_{0}$)
and regression coefficient $(\beta_{1})$ of the intensity function
(Eq. 1) that determines the relationship between elevation and the
abundance of Dickcissel (Fig. 3 a) at the Konza Prairie Biological
Station. Also, reported is Akaike information criterion (AIC) values
for four different models. 

\begin{center}
\begin{tabular*}{1.3\textwidth}{@{\extracolsep{\fill}}>{\raggedright}p{0.05\textwidth}>{\raggedright}p{0.4\textwidth}>{\raggedright}p{0.25\textwidth}>{\raggedright}p{0.22\textwidth}>{\raggedright}p{0.07\textwidth}}
\hline 
Model & Description & $\hat{\beta_{0}}$ & $\hat{\beta_{1}}$  & AIC\tabularnewline
\hline 
a & Eq. 3 (average elevation) & -9.13 (-9.44 -8.83) & 0.033 (0.022, 0.044) & 1405\tabularnewline
b & Eq. 4 (exact distances) & -9.29 (-9.66, -8.93 ) & 0.042 (0.028, 0.056) & 593\tabularnewline
c & Eq. 7 (normal distance errors) & -9.29 (-9.66, -8.93) & 0.042 (0.028, 0.056) & 594\tabularnewline
d & Eq. 7 (Laplace distance errors) & -9.29 (-9.66, -8.93) & 0.042 (0.028, 0.056) & 594\tabularnewline
\hline 
\end{tabular*}
\par\end{center}

\end{landscape}

\end{flushleft}
\end{document}